\documentclass{cjaa}                    

\usepackage{graphicx}                   

\twocolumn

\begin{document}

    \title{Gamma-Ray Spectral Characteristics of Thermal and 
Non-Thermal Emission from Three Black Holes}

    \author{James C. Ling
       \inst{1}\mailto{}
    \and William A. Wheaton
       \inst{2}}
    \offprints{J. C. Ling}                   

    \institute{Jet Propulsion Laboratory 169-327, California Institute 
of Technology,4800 Oak Grove Drive, Pasadena, CA 91109, USA\\
              \email{james.c.ling@jpl.nasa.gov}
         \and
  Infrared Processing and Analysis Center, California Institute
of Technology, 100-22, Pasadena, CA 91125, USA \\
           }

    \date{Received~~2004 July 15; accepted~~2004~~month day}

    \abstract{
Cygnus X-1 and the gamma-ray transients GRO J0422+32 and GRO 
J$1719-24$ displayed similar spectral properties when they underwent 
transitions between the high and low gamma-ray (30 keV to few MeV) 
intensity states. When these sources were in the high $\gamma$-ray 
intensity state ($\gamma_{2}$, for Cygnus X-1), their spectra 
featured two components: a Comptonized shape below 200--300 keV with a 
soft power-law tail (photon index $>$3) that extended to $\sim$1 MeV 
or beyond.  When the sources were in the low-intensity state 
($\gamma_{0}$, for Cygnus X-1), the Comptonized spectral shape below 
200 keVtypically vanished and the entire spectrum from 30 keV to 
$\sim$1 MeV can be characterized by a single power law with a 
relatively harder photon index $\sim2-2.7$. Consequently the high- 
and low-intensity gamma-ray spectra intersect, generally in the 
$\sim$400 keV - $\sim$1 MeV range, in contrast to the spectral 
pivoting seen previously at lower ($\sim$10 keV) energies. The 
presence of the power-law component in both the high- and 
low-intensity gamma-ray spectra strongly suggests that the 
non-thermal process is likely to be at work in both the high and the 
low-intensity situations. We have suggested a possible scenario (Ling 
$\&$ Wheaton, 2003), by combining the ADAF model of Esin et al. 
(1998) with a separate jet region that produces the non-thermal 
gamma-ray emission, and which explains the state transitions. Such a 
scenario will be discussed in the context of the observational 
evidence, summarized above, from the database produced by EBOP, JPL's 
BATSE earth occultation analysis system.
    }
    \authorrunning{J. C. Ling $\&$ Wm. A. Wheaton } 
    \titlerunning{Gamma-Ray Spectral Characteristics of Three Black 
Holes }  
    \maketitle
%
%
\section{Introduction}           
auto-capitalized
\label{sect:intro}
We have recently completed the processing of the full 9-year BATSE 
(Fishman et al. 1989) earth-occultation database, covering the period 
between 1991 and 2000, using the JPL Enhanced Occultation Package 
(EBOP, Ling et al.1996, 2000).
\begin{figure*}
\centering
\hspace{5mm}
\includegraphics[width=16cm, height=9cm]{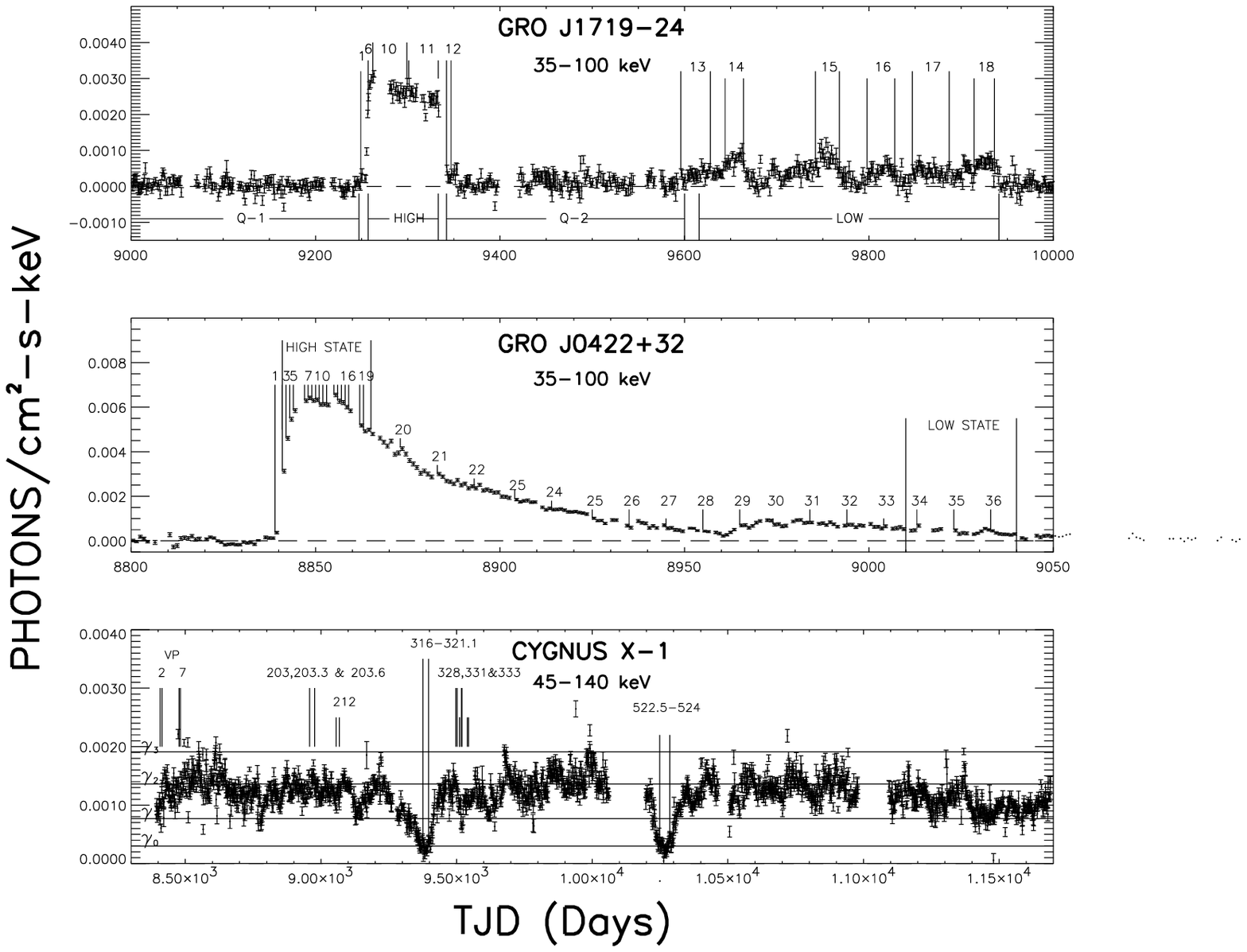}
    \caption{Soft Gamma-Ray Flux Histories of GRO J$1719-24$, GRO 
J0422+32 and Cygnus X-1  }
    \label{Fig:Flux Histories}
\end{figure*}
\begin{figure}
\centering
 
\includegraphics[width=6.7cm, height=20.5cm]{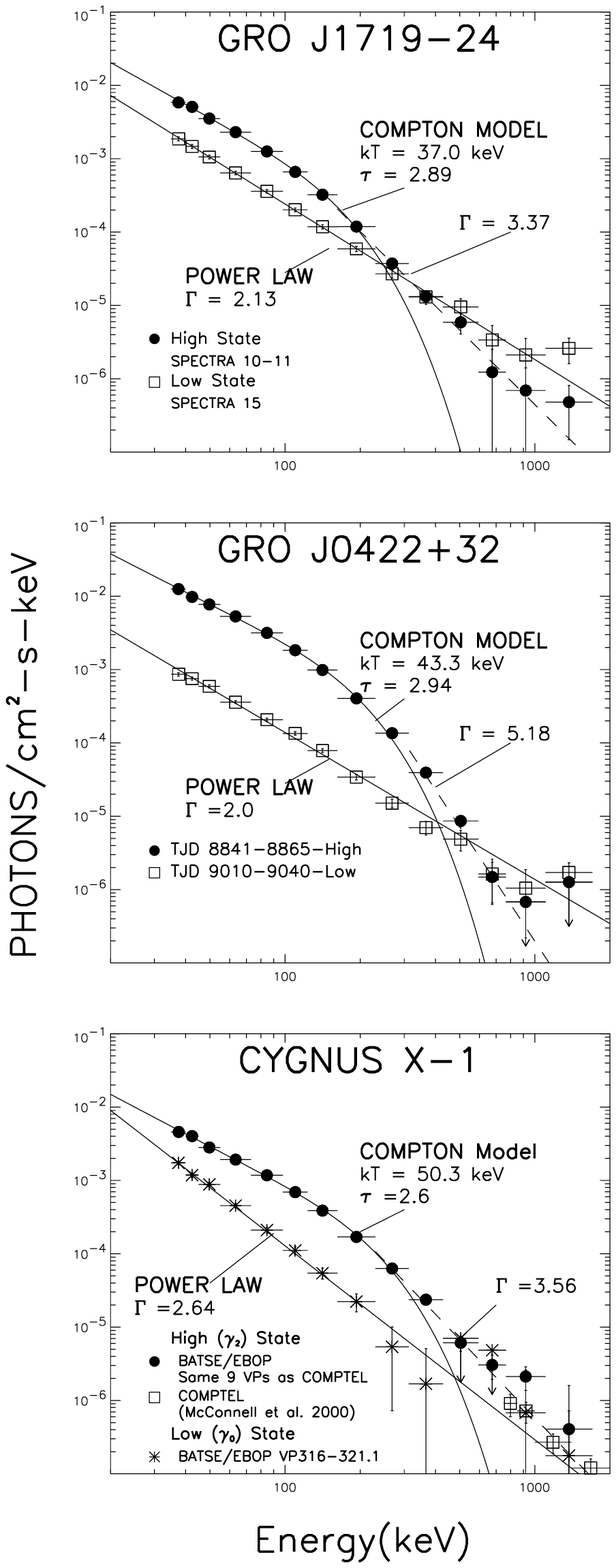}
    \caption{Comparison of High vs. Low-Intensity Gamma-Ray State Spectra }
    \label{Fig:lightcurve-ADAri}
\end{figure}
This high-level database (levels 2 $\&$ 3) consists of daily 
gamma-ray spectra with 14 energy channels between 35 keV and 1.7 MeV, 
and six broad-band flux histories with 1-day resolution for each of 
the 75 gamma-ray sources, currently included in the EBOP source 
catalog. In this paper, we focus on the soft $\gamma$-ray emission of 
three black holes in our galaxy: GRO J$1719-24$ (Ling $\&$ Wheaton 
2004), GRO J0422+32 (Ling $\&$ Wheaton 2003), and Cygnus X-1 (Ling et 
al. 1997; Zhang et al. 1997) observed by BATSE.  All three sources 
underwent multiple gamma-ray flux transitions during this 9-year 
period between low and high-intensity gamma-ray states. During these 
transitions, all three sources displayed similar spectral properties 
and variability as described in the Abstract. Section 2 presents the 
long-term soft $\gamma$-ray flux histories and spectra of these 
sources observed by BATSE. Implications of these results in enhancing 
our understanding of these systems are discussed in Section 3.

\section{Results}
\label{sect:Obs}
Figure 1 shows the soft $\gamma$-ray flux histories of GRO J$1719-24$ 
(Ling $\&$ Wheaton 2004), GRO J0422+32 (Ling $\&$ Wheaton 2003) and 
Cygnus X-1 covering the periods from 13 January 1993 to 10 October 
1995 (TJD 9000--10000), 27 June 1992 to 23 April 1993 (TJD 8800-- 
9100), and 24 May 1991 to 26 May 2000 (TJD 8400--11690, the full 
9-year CGRO mission), respectively. During these periods, all three 
sources underwent multiple flux transitions.

GRO J$1719-24$ is a low-mass X-ray binary (LMXB) with a periodicity 
of 14.7 hours. The system consists of a $\sim$4.9 solar mass compact 
object and a $\sim$1.6 solar mass companion star (Masetti et al., 
1996). The distance was estimated to be $\sim2-2.8$ kpc. The source 
was first discovered by BATSE (Harmon et al. 1993) and SIGMA (Ballet 
et al. 1993). Ling $\&$ Wheaton (2004) showed that the hard X-ray 
flux rose suddenly from the quiescent level before 17 September 1993 
(TJD 9247) to $\sim$1.5 Crab 16 days later on 3 October (TJD 9263). 
It then slowly decreased in the next 70 days to $\sim$1 Crab level on 
12 December (TJD 9333) (labeled ``10" and ``11" in Figure 1 panel 1) 
before returning roughly to the pre-transition quiescent level where 
it remained until 5 September 1994 (TJD 9600). During the next 400 
days between the fall of 1994 to the fall of 1995 (TJD 9600--10000), 
the source flared again five different times (marked as ``14", ``15", 
``16", ``17", and ``18" in Figure 1) to levels 3--7 times lower than the 
1993 high-state level. Ling $\&$ Wheaton (2004) showed that during 
the high-intensity period between TJD 9262 and TJD 9333, BATSE 
observed high-energy gamma-ray fluxes up to $\sim400-700$ keV. The 
average fluxes in the 35--100 keV, 100--200 keV, 200--300 keV 300--400 
keV, 400--700 keV and 700--1000 keV bands were measured to be (2.57 
$\pm$ 0.02) $\cdot 10^{-3}$ ($\sim$1.2 Crab),  (3.23 $\pm$ 0.05) 
$\cdot 10^{-4}$ ($\sim$1.05 Crab), (5.98 $\pm$ 0.30) $\cdot 10^{-5}$ 
($\sim$0.65 Crab), (1.65 $\pm$ 0.23) $\cdot 10^{-5}$ ($\sim$0.44 
Crab), (4.12 $\pm$ 1.29) $\cdot 10^{-6}$ ($\sim$0.28 Crab), and (1.00 
$\pm$ 0.89) $\cdot 10^{-6}$ ($\sim$0.18 Crab) photons 
cm$^{-2}$-s$^{-1}$-keV$^{-1}$, respectively. During this 1000-day 
$\gamma$-ray flux excursion, Ling $\&$ Wheaton (2004) showed that the 
source spectrum was best characterized by a power law during the 
first part of the rising phase from TJD 9249--9256. Then on TJD 9257, 
a day later, the spectrum changed suddenly from a power law to a 
Comptonized shape (Sunyaev $\&$ Titarchuk 1980) below 200 keV. It 
remained in this same Comptonized shape for the rest of the rising 
phase from TJD 9257 to TJD 9260 and when the source was in the high 
intensity state from TJD 9262 to TJD 9333. The spectrum then changed 
back to the power law for the rest of the 1000-day period when the 
source returned to the quiescent state and when the source made 
multiple transitions to the low-intensity states. Figure 2 (1st 
panel) shows that the average high-intensity state spectrum 
integrated over periods 10 $\&$ 11 has two components: a soft 
power-law tail of photon index of $3.37^{+0.19}_{-0.16}$ above 200 keV, 
that extended to $\sim$500 keV, superposed on the low-energy 
Comptonized component below 200 keV. This is compared to a typical 
low-intensity power-law spectrum with a photon index of 2.13 $\pm$ 
0.05 measured in period 15. The two spectra intercept at $\sim$400 
keV.

GRO J0422+32 is also a LMXB with a periodicity of 5.6 hours. The mass 
of the compact object is not well determined. Depending on the 
orbital inclination, a lower limit of $\geq 9 M_\odot$ is estimated. 
The distance of the source was estimated to be $\sim$2.4 kpc. The 
source was also discovered by BATSE in the summer of1992 (Paciesas et 
al. 1992) when its hard X-ray emission underwent a major outburst. 
Ling $\&$ Wheaton (2003) showed that the 35--200 keV flux rose 
sharply after the onset of the outburst on 5 August 1992 (TJD 8839, 
labeled ``1"), and reached the first of two maxima during the peak of 
the outburst on 14 August (TJD 8848, labeled ``7"). It then decreased 
slightly ($\sim5$\%
for 35--100 keV, and $\sim6$\% for 100--200 keV) over the next few days
before rising again to a second maximum on 21 August (TJD 8855,
13th tickmark). The declining phase of the event took $\sim180$ days. 
The combined 35--200 keV flux reached half the peak level on 
$\sim$TJD 8877, and a tenth of the peak on $\sim$TJD 8935, 
approximately 80 days later. It continued to decrease to a level 
about 4\% of the peak value on TJD 8960 before slowly rising again to
the so-called "secondary maximum" on TJD 8972 ($\sim120$ days after the
primary maximum on $\sim$TJD 8852) at a level about 15\% of the peak
value. The 35--200 keV flux stayed within 20\% of the ``secondary
maximum" for $\sim20$ days before declining, and finally reached the
``pre-outburst" quiescent level on TJD 9040. During this period, Ling 
$\&$ Wheaton (2003) showed that the spectrum evolved from a power law 
on TJD 8839 (marked as ``1") to a Comptonized shape below 200 keV two 
days later on TJD 8841 (2nd tick mark) where it remained until 
$\sim$TJD 8935 (marked as ``26") before it changed back to a power law 
where it remained until the end of the period. The high-intensity 
spectrum, averaged over the period from TJD 8841-8865, shows two 
components: a Comptonized shape below 200 keV followed by a soft 
power-law tail with photon index of $\sim$5.2 that extends to 
$\sim$700 keV.  The low-intensity spectrum, averaged over TJD 
$9010-9040$, is best described by a power law with photon index of 
$\sim$2. The two spectra intersect at $\sim$600 keV.

Cygnus X-1, the best-known galactic black hole, is a high mass X-ray 
binary (HMXB) with a 5.6-day periodicity. It is well known for its 
bimodal X-ray ($1-10$ keV) emission states. During an X-ray low to 
high-state transition, the hard X-ray flux ($>$10 keV) was shown to 
be anti-correlated with the soft X-ray ($<$10 keV) flux (Ling et al. 
1987; Zhang et al. 1997). Figure 1 (3rd panel) shows the $45-140$ keV 
lightcurve measured by BATSE covering the full nine years CGRO 
mission from 1991 to 2000. Included in the figure are the four flux 
levels ($\gamma_{0}, \gamma_{1}, \gamma_{2}$, \& $\gamma_{3}$) 
defined by Ling et al. (1987, 1997). The source stayed at 
approximately the $\gamma_{2}$ state during most of this period. 
However, in two occasions, around TJD $9376-9397$ (24 January--14 
February 1994; labeled as VP $316-321.1$) and TJD $10249-10286$ (15 
June--22 July 1996; VP $522.5-524$), respectively, the source made a 
transition to the $\gamma_{0}$ state. The latter was clearly 
associated with the 1996 X-ray low-to-high state transition (Zhang et 
al. 1997), while the former was believed to be also (Ling et al 1997) 
although no contemporaneous X-ray observations were available. Marked 
also in this figures are the nine CGRO Viewing Periods (VPs 2, 7, 
203, 203.3, 203.6, 212, 328, 331, \& 333) when contemporaneous 
$\gamma_{2}$ spectra measured by COMPTEL, OSSE and BATSE-EBOP were 
compared (McConnell et al. 2000). Figure 2 (3rd panel) shows a 
comparison of the $\gamma_{2}$ spectrum measured by BATSE-EBOP and 
COMPTEL for the same period with the first $\gamma_{0}$ spectrum 
measured by BATSE in 1994. As previously reported in the literature 
(Ling et al. 1997; McConnell et al. 2000), the $\gamma_{2}$ spectrum 
shows two components, a Comptonized shape below 300 keV followed by a 
power law from 300  keV to $\sim$1 MeV. The $\gamma_{0}$ spectrum 
(Ling et al. 1997; Phlipset al. 1996), on the other hand, shows a 
single power law with photon index of $\sim2.6-2.7$. The two spectra 
intersect at $\sim$1 MeV. McConnell et al. (2002) reported that the 
same $\gamma_{2}$ spectrum intersects the 2nd $\gamma_{0}$ spectrum 
measured in 1996 at $\sim$1 MeV also.

\begin{figure*}
\centering
\includegraphics[width=15cm, height=8.5cm]{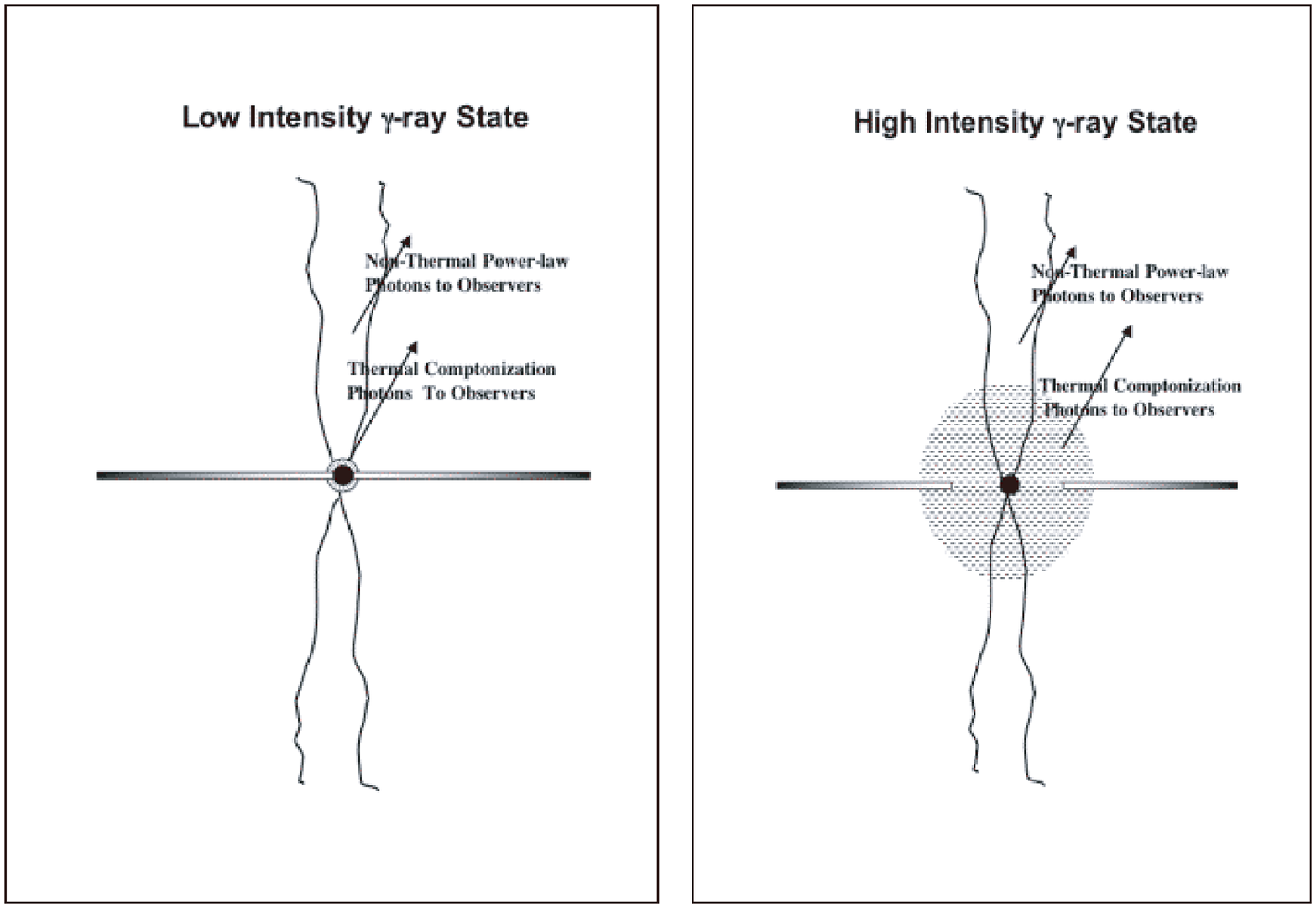}
    \caption{System Configurations for High and Low Gamma-Ray 
Intensity States   }
    \label{Fig:System Configuration}
\end{figure*}

\section{Discussion}
\label{sect:discussion}
GRO J$1719-24$ (Ling $\&$ Wheaton 2004), GRO J0422+32 (Ling $\&$ 
Wheaton 2003) and Cygnus X-1 (Philips et al, 1996; Ling et al. 1997; 
McConnell et al. 2000, 2002) are three galactic black holes that 
displayed similar gamma-ray spectral characteristics when undergoing 
transitions between the high and low $\gamma$-ray intensity states. 
The low-intensity states were described by a power law with spectral 
index of $\sim2-2.7$. The high-intensity states were well fit by a 
two-component spectrum: a Comptonized component below $\sim$200 keV 
with electron temperature $kT_e$ $\sim37-50$ keV and optical depth 
$\tau$ $\sim2.6-3$, and a soft power-law tail with photon index of 
$>$3.1 above 200 keV. These two distinct gamma-ray state spectra are 
clearly different than the simple power law spectrum shown in GRO 
J1655--40, and a broken power law shown in GRS1915+105 (Case et al. 
2004), two well-known microquasars in our galaxy, suggesting that 
there are perhaps two different classes of gamma-ray emitting BH 
candidates in our galaxy.

The persistent power-law emission seen in both high and low-intensity 
spectra of these sources suggests that non-thermal emission may be at 
work in both situation. Non-thermal emission may be associated with 
jets which Meier (2001) claimed is a natural consequence of accretion 
flows onto rotating black holes. More recently, Meier's (2004) model 
of Magnetically-Dominated Accretion Flows (MDAF) further develops the 
idea and integrates accretion and jet production as an extension of 
the ADAF (Esin et al. 1998) and Shakura $\&$ Sunyaev (1976) disk 
models. Non thermal emission may also be associated with Compton 
scattering off relativistic electrons of free infalling matter onto 
the black hole in the converging flow region near the event horizon 
(Chakrabarti $\&$ Titarchuk 1995; Turolla et al. 2002). Over the 
years. there have been also attempts to develop a hybrid 
thermal/non-thermal Comptonization model (Coppi 1998;  Gierlinski et 
al.1999) to explain the two-component high-intensity gamma-ray 
spectra observed in BH binary systems such as GRO J$1719-24$, GRO 
J0422+32 and Cygnus X-1, but no attempts have been made to explain 
how the spectrum evolved into a power law in the low-intensity 
scenario. Ling $\&$ Wheaton (2003) suggested a possible system 
scenario for explaining both the high and low-intensity spectra (see 
Figure 3). It is based on the ADAF model of Esin et al (1998) along 
with the source geometry envisioned by Poutanen $\&$ Coppi (1998) and 
others. For the high-intensity scenario (Figure 3 right panel), the 
system consists of a hot inner corona, a cooler outer thin disk, and 
a separate region that produces the power-law $\gamma$-ray emission. 
We hereby refer to the latter the "Non Thermal Emission Region 
(NTER)". NTER may include a jet (see Figure 3) and possibly also the 
converging flow region discussed above. Under such condition, the 
transition radius of the disk is $\sim$100 Schwarzschild radii from 
the black hole. Electrons in the hot corona up-scattered the 
low-energy photons produced both inside the corona as well as from 
the outer disk to form the Comptonized component that dominates the 
spectrum in the 35--200 keV range. These same electrons also 
down-scattered the high energy photons ($>$10 MeV) produced in NTER 
resulted in forming a softer power-law component observed in the 300 
keV to 1 MeV range. Under the low-intensity scenario that could be 
triggered by a significant increase of the accretion rate, a large 
quantity of soft photon flux was produced in the disk that cooled and 
quenched the hot corona and movedthe transition radius inward to a 
distance very close to the horizon. Under this condition, the 
Comptonized component below 200 keV disappeared and the entire 
35--1000 keV spectrum is dominated by the unperturbed power-law 
emission produced in NTER.

\begin{acknowledgements}
We wish to thank Gerald Fishman and his BATSE team for their support 
of the JPL EBOP development throughout the years, and Michael Cherry 
and Gary Case for their comments on this paper. The work described in 
this paper was carried out at the Jet Propulsion Laboratory under the 
contract with the National Aeronautics and Space Administration.
\end{acknowledgements}

\label{lastpage}

\end{document}